\documentclass[a4paper,11pt]{article}
\usepackage{pos}

\title{Formation and Decay of Oscillons in Einstein–Cartan Higgs Inflation}

\author*[a]{Javier Rubio}

\affiliation[a]{Departamento de Física Teórica and Instituto de Física de Partículas y del Cosmos (IPARCOS-UCM), Universidad Complutense de Madrid, 28040 
Madrid, Spain} 

\emailAdd{javier.rubio@ucm.es}

\abstract{We review recent progress in the understanding of the preheating stage of Higgs inflation formulated within the Einstein–Cartan framework of gravity. This setup  smoothly interpolates between the metric and Palatini formulations of the theory, leading to a distinctive phenomenology in an intermediate regime. Following the end of inflation, the Higgs field undergoes a non-trivial out-of-equilibrium evolution driven by tachyonic instabilities and nonlinear self-interactions, which fragment the inflaton condensate and give rise to well-localized oscillon configurations. While early studies suggested the formation of long-lived oscillons and the possibility of an extended matter-dominated phase, more recent analyses show that self-interactions at small field values render these objects transient, eventually triggering their decay and the onset of radiation domination. We discuss the implications of this dynamics for the thermal history of the Universe, the inflationary observables, and the generation of stochastic gravitational waves.}
\FullConference{
}

\begin{document}
\maketitle

\section{Introduction}

Cosmic inflation provides a compelling explanation for the observed homogeneity and isotropy of the Universe and for the near scale invariance of primordial density perturbations, while simultaneously offering a dynamical origin for the initial conditions of the hot Big Bang \cite{Guth:1980zm, Linde:1981mu, Mukhanov:1981xt}. Despite its phenomenological success, the particle-physics nature of the inflaton field and the details of the transition from inflation to a radiation-dominated Universe remain open. In particular, a predictive inflationary framework must come equipped with a well-defined mechanism for transferring the energy stored in the homogeneous inflaton condensate into relativistic degrees of freedom \cite{Barman:2025lvk}, with significant differences between oscillatory \cite{Kofman:1994rk,Kofman:1997yn,Greene:1997fu,Felder:2000hj,Copeland:2002ku,Felder:2001kt,Barman:2023opy} and non-oscillatory settings \cite{Bernal:2020bfj,Bettoni:2021qfs,Bettoni:2019dcw,Laverda:2023uqv}. 

Higgs inflation (HI) stands out among inflationary scenarios for its minimality: the Standard Model Higgs field itself can play the role of the inflaton through a non-minimal coupling to gravity \cite{Bezrukov:2007ep,Rubio:2018ogq}. This possibility is especially attractive because the relevant couplings are, in principle, known (or at least constrained) at high energies \cite{Bezrukov:2009db,Bezrukov:2012sa,Degrassi:2012ry,Buttazzo:2013uya}, making the post-inflationary dynamics potentially computable \cite{Bezrukov:2014bra,Hamada:2014wna, Bezrukov:2014ipa,Rasanen:2017ivk}.

At the same time, HI is known to be sensitive to the precise formulation of gravity and the associated higher-dimensional operator structure induced at large field values \cite{Bezrukov:2010jz,Shaposhnikov:2020fdv,Shaposhnikov:2020gts,Shaposhnikov:2025znm}. As a consequence, the efficiency of preheating and the duration of the heating stage can vary significantly across theories that are otherwise very similar during slow roll \cite{ Garcia-Bellido:2008ycs,Bezrukov:2008ut,Rubio:2015zia,Repond:2016sol,Rubio:2019ypq,Piani:2023aof,Piani:2025dpy}. Since the post-inflationary expansion history affects the mapping between inflationary dynamics and late-time observables, understanding reheating becomes essential for precision cosmology.

The heating epoch is inherently nonlinear. Following the end of inflation, the homogeneous condensate typically undergoes a phase of non-perturbative particle production, driven by parametric resonance and/or tachyonic instabilities, which can lead to fragmentation of the inflaton condensate \cite{Barman:2025lvk}. In theories with suitable potentials, this process can generate long-lived, spatially localized field configurations known as oscillons \cite{Bogolyubsky:1976yu,Gleiser:1993pt,Copeland:1995fq,Amin:2010dc,Amin:2011hj,Antusch:2015nla,Lozanov:2017hjm,Hasegawa:2017iay,Antusch:2017flz,Sang:2019ndv,Antusch:2019qrr,Ibe:2019lzv,Kou:2019bbc,Sang:2020kpd,Aurrekoetxea:2023jwd,Mahbub:2023faw,vanDissel:2023zva}. Such objects can temporarily behave as pressureless matter, modify the pre-radiation expansion history, and source stochastic gravitational waves through their anisotropic stresses \cite{Zhou:2013tsa,Antusch:2016con,Antusch:2017vga,Amin:2018xfe,Liu:2018rrt,Lozanov:2019ylm,Lozanov:2022yoy,Piani:2023aof}. These effects are especially relevant in scenarios where the potential admits a broad region that is shallower than quadratic, allowing localized lumps to persist for many oscillations.

In this contribution, we focus on HI embedded in Einstein--Cartan (EC) gravity, a formulation in which spacetime torsion is allowed but does not introduce additional propagating gravitational degrees of freedom \cite{Utiyama:1956sy,Kibble:1961ba}. After integrating out this new component, the framework yields a predictive scalar-tensor dynamics in which the inflationary potential and kinetic structure are determined by gravitational interactions. This setup leads to a distinctive post-inflationary phenomenology: tachyonic amplification can efficiently trigger fragmentation and oscillon formation in an intermediate regime between the standard metric ~\cite{Bezrukov:2007ep,Barbon:2009ya,Bezrukov:2009db,Burgess:2010zq,Bezrukov:2010jz,Giudice:2010ka,Bezrukov:2014bra,Hamada:2014iga,Hamada:2014wna,George:2015nza,Fumagalli:2016lls,Bezrukov:2017dyv} and Palatini limits ~\cite{Bauer:2008zj,Bauer:2010jg,Rasanen:2017ivk,Enckell:2018kkc,Rasanen:2018fom,Rasanen:2018ihz,Gialamas:2019nly,Rubio:2019ypq,Shaposhnikov:2020fdv,Annala:2021zdt,Dux:2022kuk,Poisson:2023tja} of the theory. A key point to be emphasized here is that the fate of these oscillons is controlled by the usual Higgs-like structure of the potential at small field values. In particular, the presence of a quartic regime generically renders oscillons short-lived, and their decay drives a comparatively rapid onset of radiation domination, bounding the duration of the heating stage and stabilizing the associated predictions for inflationary observables.

The rest of these proceedings is organized as follows. In Section 2 we introduce the Einstein–Cartan Higgs inflation framework and its effective description. Section 3 is devoted to the post-inflationary dynamics and formation of oscillons. In Section 4 we analyze their evolution and eventual decay, while in Section 5 we discuss the resulting cosmological implications. Finally, we summarize our findings in Section 6.

\section{Higgs inflation in Einstein--Cartan gravity}

\begin{figure}
  \centering
  \includegraphics[width=0.8\textwidth]{\detokenize{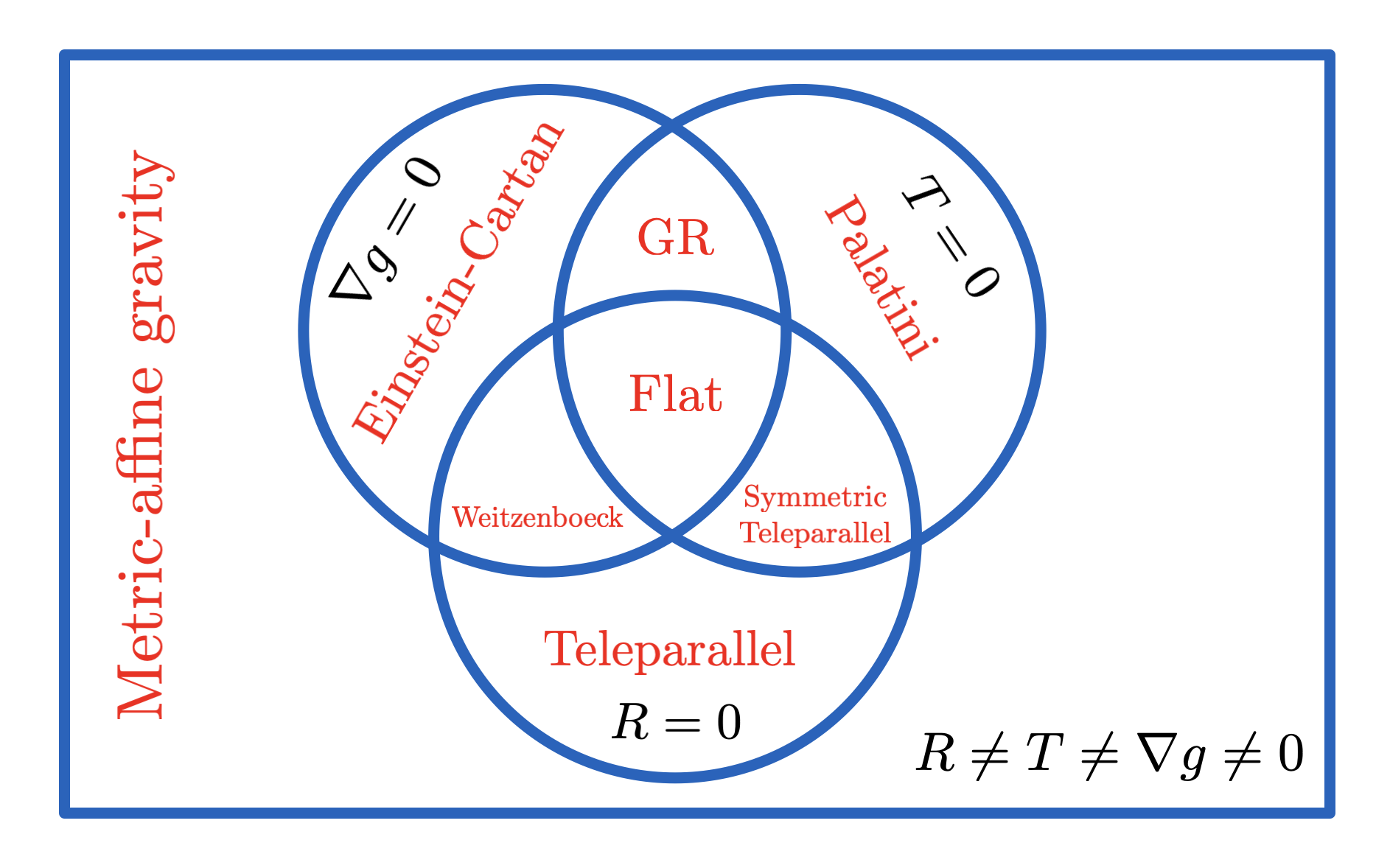}}
  \caption{Schematic map of metric--affine formulations of gravity in terms of non-metricity ($\nabla g\neq 0$), torsion ($T\neq 0$) and curvature ($R\neq 0$). GR sits at the intersection of vanishing torsion and non-metricity, while Palatini and EC correspond to different restrictions on $(\nabla g,\,T,\,R)$. Teleparallel and symmetric-teleparallel descriptions trade curvature for torsion or non-metricity, respectively.}
  \label{fig:metric_affine_venn}
\end{figure}

A useful way to organize possible ``formulations'' of gravity is to distinguish the metric as the field that measures distances from the connection as the field that defines parallel transport. This separation provides a convenient classification of gravitational theories in terms of curvature, torsion and non-metricity. In particular, metric–affine formulations can be characterized by which of these geometrical quantities are allowed to be non-vanishing, as schematically illustrated in Fig.~\ref{fig:metric_affine_venn}. In this language, General Relativity (GR) corresponds to choosing the Levi--Civita connection, which is torsion-free and metric-compatible. EC gravity generalizes this by allowing for a non-vanishing torsion tensor,
\begin{equation}
T^\rho{}_{\mu\nu}\equiv \Gamma^\rho_{\mu\nu}-\Gamma^\rho_{\nu\mu}\,,
\end{equation}
while preserving local Lorentz invariance. For the class of actions relevant here, torsion does not introduce additional propagating gravitational degrees of freedom. Instead, it appears algebraically and can be eliminated through its equations of motion. The resulting theory is therefore dynamically equivalent to GR in the pure gravitational sector, containing only the massless graviton. However, integrating out torsion generates specific interactions in the matter sector, particularly for fermions and for scalars non-minimally coupled to curvature \cite{Shaposhnikov:2020frq,Karananas:2021zkl}. In the context of Higgs inflation, this procedure leads to an effective scalar--tensor theory in the Einstein frame with a characteristic non-canonical kinetic structure, which we construct explicitly in what follows.

\subsection{Einstein-frame action and constant-\texorpdfstring{$c$}{c} description}

We restrict our analysis to Jordan-frame EC realizations of HI constructed from polynomial operators with canonical dimension $d\leq 4$ and containing at most two derivatives acting on the fields \cite{Langvik:2020nrs,Shaposhnikov:2020gts,Karananas:2021zkl}. The restriction to $d\leq 4$ guarantees that any non-renormalizable contributions originate solely from the gravitational sector, thus maintaining compatibility with the effective field theory (EFT) interpretation of GR. Limiting the action to operators with no more than two derivatives avoids the appearance of extra propagating modes beyond the Higgs field and the two transverse graviton polarizations.

Under these assumptions, the most general local and Lorentz-invariant action considered in this work can be written as
\begin{equation}\label{S-EH}
S[g, h, T^2,\partial T]  =\int d^4x\sqrt{- g} \left[ \frac{ f(h)}{2} R-\frac{1}{2}  g^{\mu\nu}\partial_\mu h \partial_\nu h- U(h)\right]+S_T[T^2,\partial T]\,,
\end{equation}
with \(h\) denoting the Higgs field in the unitary gauge,
\begin{equation}
\label{eq:function-HINY0}
f(h)=1+\xi h^2 \,,
\hspace{15mm}
U(h)\simeq \frac{\lambda}{4}   h^4\,,
\end{equation}
and $\xi$ and $\lambda$ dimensionless parameters. Here we have neglected the electroweak vacuum expectation value of the Higgs field, \(v_{\rm EW}\simeq 246\,{\rm GeV}\), since the inflationary dynamics occurs at field values \(h\gg v_{\rm EW}\), rendering the quadratic and constant terms in the potential irrelevant for the regimes considered in this work. The term $S_T$ collects all admissible torsion operators \cite{Langvik:2020nrs,Shaposhnikov:2020gts,Karananas:2021zkl},
\begin{equation}
\begin{aligned}
\label{eq:action_torsion}
S_{\rm T} &= \int d^4 x \,\sqrt{-g}\Bigg[ v^\mu \partial_\mu Z^v + a^\mu \partial_\mu Z^a \\
&+ \frac{1}{2}\Big(G_{vv} v_\mu v^\mu  + 2G_{va}v_\mu a^\mu + G_{aa}	a_\mu a^\mu 
 +G_{\tau\tau}\tau_{\alpha\beta\gamma} \tau^{\alpha\beta\gamma}+ \tilde{G}_{\tau\tau} \epsilon^{\mu \nu \rho \sigma} \tau_{\lambda\mu\nu} \tau^\lambda_{~\rho\sigma}\Big) \Bigg]\,,
\end{aligned}
\end{equation}
with
\begin{equation}
Z^{v/a}=\zeta^{v/a}_h h^2 \,,  \hspace{15mm}
G_{ij}= c_{ij} \left(1+\xi_{ij}h^2\right) \,,
\end{equation}
no summation over repeated $i,j$ indices implied and $\zeta^{v/a}_h$, $c_{ij}$, and $\xi_{ij}$ constant coefficients. The quantities
\begin{equation}
v_\mu = T^\nu_{~\mu\nu} \,, \hspace{10mm}
a_\mu = \epsilon_{\mu\nu\rho\sigma}T^{\nu\rho\sigma} \,,
\hspace{10mm}
\tau_{\mu\nu\rho} =\frac 2 3 \left( T_{\mu\nu\rho} -v_{[\nu} g_{\rho]\mu} - T_{[\nu\rho]\mu} \right) \,,
\end{equation}
represent, respectively, the vector, axial (pseudo-vector), and traceless tensor irreducible components of the torsion tensor, which decomposes as
\begin{equation}
\label{eq:tors_irreps}
T_{\mu\nu\rho} = e_{\mu A} T^A_{\nu\rho}= \frac23 v_{[\nu}g_{\rho]\mu} - \frac16 a^\sigma \epsilon_{\mu\nu\rho\sigma} +\tau_{\mu\nu\rho} \,,
\end{equation}
with the square brackets denoting antisymmetrization over the enclosed indices.

In order to get a more intuitive picture,  one can perform a Weyl rescaling of the metric to the Einstein frame and eliminate the torsion components through their equations of motion. The resulting theory can be written as
\begin{equation}
S = \int d^4x \sqrt{-g}\, \left[ \frac{M_P^2}{2}R
- \frac{1}{2} K(h) (\partial h)^2- V(h) \right]\,,
\label{eq:EF_action}
\end{equation}
with
\begin{equation}
\label{eq:K-V}
K(h)=\frac{1+ c(h) \, h^2}{(1+\xi h^2)^2}\,, \hspace{15mm} V(h)=\frac{U(h)}{(1+\xi h^2)^2}\,,
\end{equation}
and
\begin{equation}
c(h) = \xi + 6\xi^2 + 4f(h) \frac{G_{aa}(\zeta^{v}_h)^2 + G_{vv}(\zeta^{a}_h)^2 - G_{va}\zeta^{v}_h\zeta^{a}_h}{G_{vv}G_{aa}-G_{va}^2}
\end{equation}
a field-dependent quantity fixed by the full set of dimensionless couplings of the theory. In this formulation, torsion leaves its imprint exclusively through a controlled subset of higher-dimensional operators in the scalar sector. The geometric origin of these terms implies non-trivial selection rules, drastically reducing the freedom typically present in a generic EFT and constraining the structure of admissible non-renormalizable interactions. Note that $G_{\tau\tau}$ and $\tilde{G}_{\tau\tau}$ are absent from the Einstein-frame kinetic structure, since the equations of motion for the tensorial irreducible component of torsion imply $\tau_{\mu\nu\lambda}=0$~\cite{Gialamas:2024iyu}.

\begin{figure}
  \centering
  \includegraphics[width=0.96\textwidth]{\detokenize{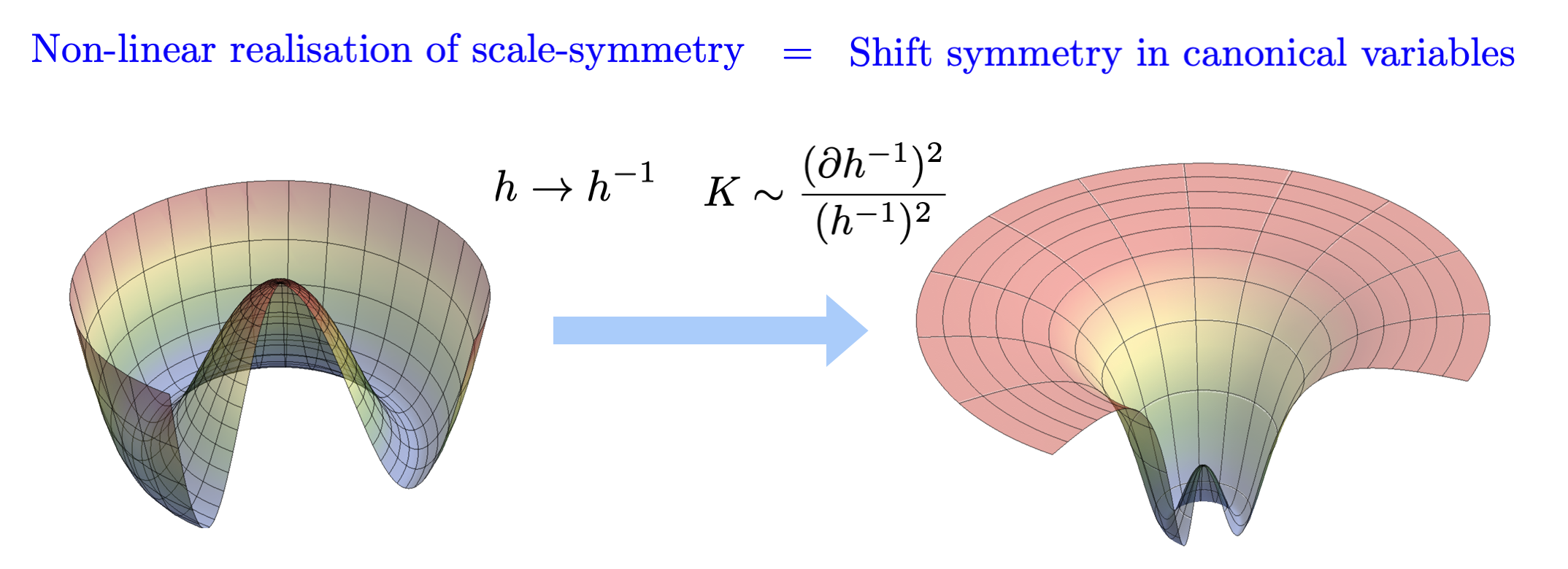}}
  \caption{Geometric intuition for the emergence of an approximate shift symmetry in suitable canonical variables. A non-linear field redefinition can turn a scale-symmetric structure in the original variables into an approximately shift-symmetric regime in the canonically normalised field. In Higgs-inflation-like models this mechanism underlies the flattening of the Einstein-frame potential at large field values and the associated attractor behaviour of inflationary observables \cite{Rubio:2018ogq}.}
  \label{fig:scale_to_shift_geometry}
\end{figure}

A sufficiently broad and phenomenologically accurate description of EC Higgs inflation can be obtained by parametrizing torsion-induced effects through an approximately constant parameter $c$ over the field range relevant for inflation and the onset of preheating \cite{Piani:2022gon,Piani:2023aof}. Choosing, for instance,
$c_{vv} = -2/3\,, \hspace{2mm} c_{va} =0\,, \hspace{2mm}   c_{aa}=1/24\,, \hspace{2mm} \xi_{vv}=\xi_{aa}=-\zeta^{v}_h=\xi\,, 
\hspace{2mm} \xi_{va}=0\,, \hspace{2mm}
\zeta^{a}_h=\xi_\eta/4,$
one obtains an effective parameter $c=\xi + 6\xi_\eta^2$, where $\xi_\eta$ is a non-minimal coupling to be constrained observationally. In this limit, the general expression in Eq.~\eqref{eq:action_torsion} simplifies to an interaction term proportional to the Nieh--Yan topological invariant~\cite{Nieh:1981ww,Nieh:2008btw},
\begin{equation}
    \label{eq:Nieh-Yan}
    S_{\rm T} =- \frac{1}{4}\int d^4x\:\xi_\eta h^2\partial_{\mu}\left(\sqrt{-g}\epsilon^{\mu\nu\rho\sigma}T_{\nu\rho\sigma}\right)~,
\end{equation} 
where $\epsilon^{\mu\nu\rho\sigma}$ denotes the totally antisymmetric tensor ($\epsilon_{0123}=1$). A different parameter choice, namely $\xi_{vv}=\xi_{aa}=\xi$ and $c_{av}=0$, also leads to a constant value 
\begin{equation}
c= \xi+  6\xi ^2+ 4\left( \frac{\zeta_{ha}^2}{c_{aa}}+ \frac{\zeta_{hv}^2}{c_{vv}}\right)\,,
\end{equation}
thus reproducing the same phenomenology discussed in this work for any set of parameters $\xi$, $\zeta_{ha}$, $\zeta_{hv}$, $c_{aa}$ and $c_{vv}$ that yields the same numerical value of $c$. Under these conditions, the Einstein-frame potential preserves the familiar Higgs-inflation plateau at large $h$ values, while the kinetic structure interpolates between distinct effective ``universality classes'' of scalar dynamics. Crucially, the parameter $c$ controls how the canonical field $\phi$ stretches relative to the Jordan-frame Higgs field $h$, thereby determining how efficiently large-field configurations are mapped into the intermediate region. This behaviour is illustrated in Fig.~\ref{fig:scale_to_shift_geometry}, where a nonlinear field redefinition  
\begin{equation}
\frac{d\phi}{dh}=\sqrt{K(h)}
\label{eq:chi_def}
\end{equation}
is shown to convert an approximate scale symmetry in the Jordan frame into an approximate shift symmetry in the canonical variable at large field values.

In terms of the canonically normalized field \(\phi\) defined in Eq.~\eqref{eq:chi_def}, the Einstein-frame potential exhibits three characteristic regimes: quartic at small field values, an intermediate quadratic regime, and an exponentially flat plateau at large field values. 
\begin{figure}
  \centering
  \includegraphics[width=0.8\textwidth]{\detokenize{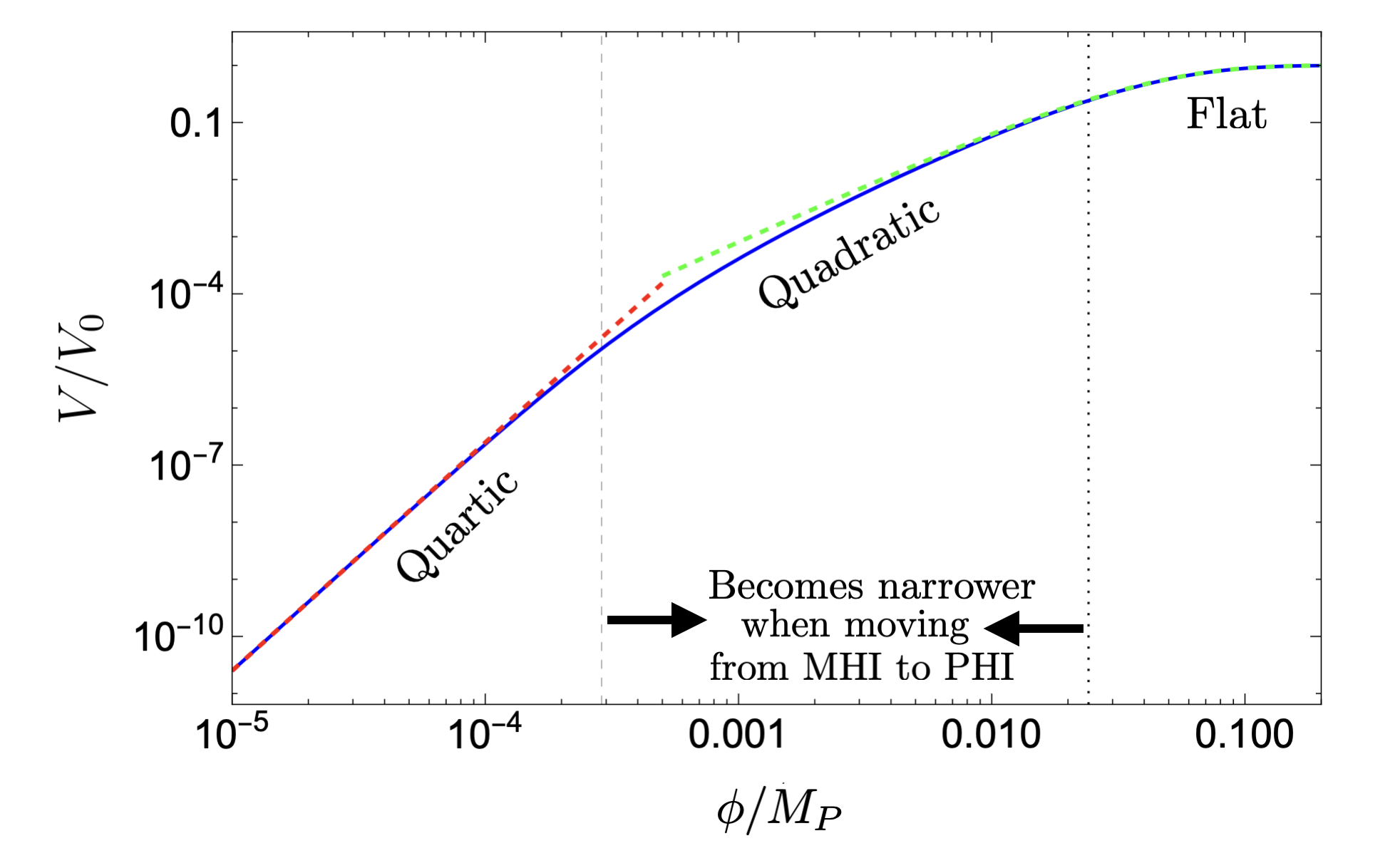}}
  \caption{Representative shape of the Einstein-frame potential across field ranges relevant for post-inflationary dynamics, with $\lambda = 0.001$, $\xi = 5 \cdot 10^4$, and $c = 1.21 \cdot 10^7$. At small amplitudes the potential is effectively quartic, at intermediate amplitudes it is approximately quadratic, and at large amplitudes it approaches a plateau. The relative extent of the intermediate quasi-quadratic regime depends on the underlying gravitational formulation, and plays a key role in determining whether the system efficiently fragments and forms oscillons during preheating. The arrows indicate that this regime becomes narrower when moving from metric Higgs inflation (MHI) to Palatini Higgs inflation (PHI).}
  \label{fig:potential_regimes}
\end{figure}
A useful approximation is, therefore,
\begin{equation}
V(\phi)\simeq
\begin{cases}
\displaystyle \frac{\lambda}{4}\phi^4, & \phi\lesssim \phi_c,\\[6pt]
\displaystyle \frac12 M^2\phi^2, & \phi_c\lesssim \phi \lesssim \phi_i\,,\\[6pt]
\displaystyle \frac{\lambda M_P^4}{4\xi^2}\left[1-\exp\!\left(-\frac{2\xi|\phi|}{\sqrt{c}\,M_P}\right)\right]^2\,,
& \phi \gtrsim \phi_i\,,
\end{cases}
\label{eq:V_piecewise}
\end{equation}
where the crossover scale $\phi_c$ and intermediate mass $M$ are respectively given by
\begin{equation}
\phi_c \equiv \frac{M_P}{\sqrt{c}}\,,
\qquad\qquad
M^2 \equiv \frac{2\lambda}{c}\,M_P^2\,,
\label{eq:chi_c_M}
\end{equation}
while the inflection point $\phi_i$ separating the plateau from the intermediate region is located at
\begin{equation}
\phi_i \equiv \frac{\sqrt{c}\,M_P}{2\xi}\ln 2\,.
\label{eq:chi_i}
\end{equation}
We stress that Eq.~\eqref{eq:V_piecewise} should be understood as an asymptotic representation of the exact Einstein-frame potential in the different field regimes. The underlying potential is a smooth and analytic function of the canonical field, and the transitions between the quartic, quadratic, and plateau regions occur continuously. The piecewise form \eqref{eq:V_piecewise} merely provides a convenient approximation that makes manifest the dominant scaling behavior in each regime, cf. Fig.~\ref{fig:potential_regimes}. Note also that the large field behaviour effectively matches that of the $\alpha$-attractor $E$-models~\cite{Kallosh:2021mnu,Artymowski:2016pjz}, upon the formal identification $c \rightarrow 6\xi^2\alpha$, with the additional feature that the theory now exhibits a discrete $\mathbb{Z}_2$ symmetry.

In the constant-\(c\) regime, the main inflationary observables following from the inflationary region $\phi\gtrsim \phi_i$ can be expressed compactly in terms of the number of e-folds \({\cal N}_*\) between horizon exit of the pivot scale and the end of inflation, 
\begin{equation}
A_s \simeq \frac{\lambda {\cal N}_*^2}{12\pi^2 c}\,,
\qquad\qquad
n_s \simeq 1-\frac{2}{{\cal N}_*}\,,
\qquad\qquad
r \simeq \frac{2c}{\xi^2{\cal N}_*^2}\,.
\label{eq:infl_obs}
\end{equation}
While the spectral tilt \(n_s\) is primarily controlled by \({\cal N}_*\), the amplitude of the curvature power spectrum $A_s$ and the tensor-to-scalar ratio \(r\) depend respectively on $c$ and the effective ratio \(c/\xi^2\). For $\xi = \sqrt{c/6}$, the resulting predictions coincide with those of metric HI models~\cite{Bezrukov:2007ep,Barbon:2009ya,Bezrukov:2009db,Burgess:2010zq,Bezrukov:2010jz,Giudice:2010ka,Bezrukov:2014bra,Hamada:2014iga,Hamada:2014wna,George:2015nza,Fumagalli:2016lls,Bezrukov:2017dyv} and related multifield and scale-invariant frameworks~\cite{Shaposhnikov:2008xb,Garcia-Bellido:2011kqb,Garcia-Bellido:2012npk,Bezrukov:2012hx,Rubio:2014wta,Karananas:2016kyt,Trashorras:2016azl,Casas:2017wjh,Tokareva:2017nng,Casas:2018fum,Shaposhnikov:2018jag,Herrero-Valea:2019hde,Almeida:2018oid,Karananas:2020qkp,Rubio:2020zht,Shaposhnikov:2020frq,Karananas:2021gco,Gialamas:2021enw,Piani:2022gon,Belokon:2022pqf,Karananas:2023zgg}. In this regime, both $n_s$ and $r$ become insensitive to the precise magnitude of $\xi$ (see~\cite{Rubio:2018ogq} for a detailed overview).
On the other hand, when $\xi=c$, the predictions match those characteristic of Palatini HI and its generalizations~\cite{Bauer:2008zj,Bauer:2010jg,Rasanen:2017ivk,Enckell:2018kkc,Rasanen:2018fom,Rasanen:2018ihz,Gialamas:2019nly,Rubio:2019ypq,Shaposhnikov:2020fdv,Annala:2021zdt,Dux:2022kuk,Poisson:2023tja}, where the tensor-to-scalar ratio is parametrically suppressed by a factor $1/\xi$. This structure neatly disentangles the inflationary universality class from the post-inflationary history: for fixed \((c,\xi)\), uncertainties in reheating feed into \({\cal N}_*\), shifting \((n_s,r)\) within a narrow but non-negligible window. It is therefore important to understand whether the heating stage allows for a wide range of expansion histories or is constrained by robust dynamical mechanisms.

The hierarchy between $\phi_c$, $\phi_i$ and the inflationary field values encodes the full post-inflationary dynamics. While slow-roll inflation probes exclusively the plateau region, reheating necessarily explores the transition between the quadratic and quartic regimes. The relative width of the intermediate quadratic window controls both the number of oscillations spent in a tachyonic band and the typical amplitude at which nonlinear effects become important. In this sense, the reheating stage acts as a magnifying glass of the gravitational formulation: small differences in kinetic structure translate into parametrically different nonlinear outcomes.
\begin{figure}
  \centering
  \includegraphics[width=0.87\textwidth]{\detokenize{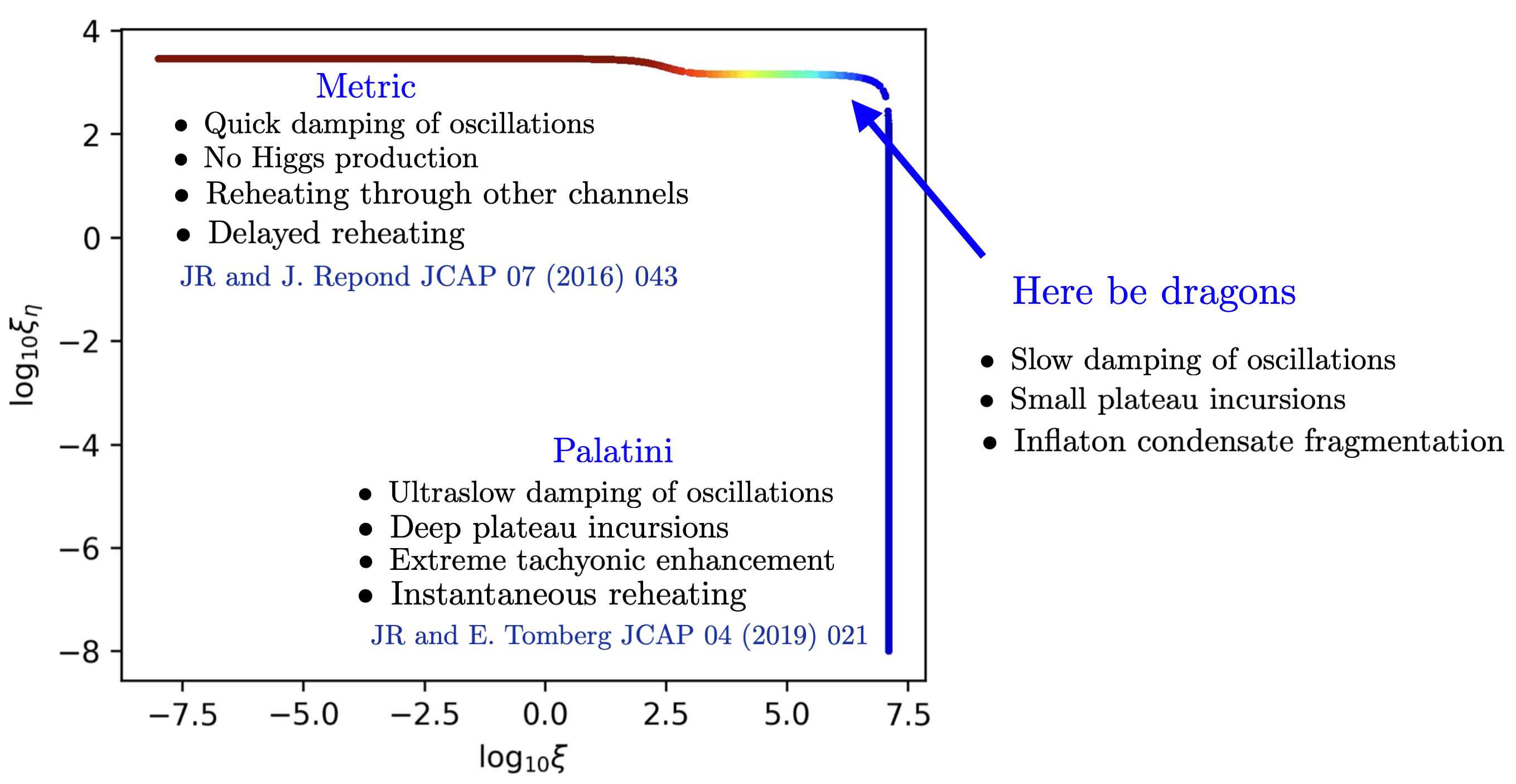}}
  \caption{Qualitative comparison of the inflationary and post-inflationary behaviours across the metric and Palatini limits and the intermediate regime for a scenario with $c=\xi+6\xi_\eta^2$. The colour scale indicates the tensor-to-scalar ratio $r$ in the displayed region. The different damping rates of the homogeneous inflaton oscillations and the typical depth of plateau incursions control the strength and repetition of tachyonic amplification episodes. In the intermediate region the inflaton condensate can fragment efficiently, providing favourable conditions for oscillon formation before backreaction and eventual decay drive the system towards radiation domination. Adapted from \cite{Shaposhnikov:2020gts}.} 
  \label{fig:parameter_space_metric_palatini_EC}
\end{figure}
In practice, one finds three broad regimes:
(i) a strongly damped regime in which the background quickly settles into small oscillations and Higgs self-production is inefficient;
(ii) an ultra-slow damping regime with large excursions and very efficient non-perturbative amplification;
and (iii) an intermediate regime in which repeated tachyonic episodes amplify fluctuations until the condensate fragments, but without an instantaneous transfer of energy into a fully radiation-like state, cf.~Fig.~\ref{fig:parameter_space_metric_palatini_EC}. The latter regime is the most relevant for oscillon formation and for a controlled, computable heating stage.

\section{Preheating and tachyonic amplification}

Immediately after inflation, the energy density of the Universe is stored predominantly in the homogeneous inflaton condensate. The subsequent heating stage is governed by the conversion of this energy into inhomogeneous excitations. At the level of the canonical field \(\phi\), the dynamics follow from the corresponding Klein-Gordon equation evaluated on a flat Friedmann-Lema\^itre-Robertson-Walker background $ds^2=dt^2-a^2(t)\delta_{ij}dx^i dx^j $ with scale factor $a(t)$,
\begin{equation}
\ddot\phi + 3H\dot\phi - \frac{1}{a^2}\nabla^2\phi + V_{,\phi}=0,¡\,,
\label{eq:KG_full}
\end{equation}
with the expansion rate $H\equiv \dot a/a$ determined by
\begin{equation}
3M_P^2 H^2 = \frac12\dot\phi^2+\frac{(\nabla\phi)^2}{2a^2}+V(\phi)\,,
\qquad
\dot H = -\frac{1}{2M_P^2}\left(\dot\phi^2+\frac{(\nabla\phi)^2}{3a^2}\right)\,.
\label{eq:Friedmann}
\end{equation}
This system admits a stage in which the background field oscillates around the minimum while fluctuations are amplified and eventually become nonlinear. To understand the onset of particle production, we decompose the field into a homogeneous background and small fluctuations, $\phi(\mathbf{x},t)=\bar\phi(t)+\delta\phi(\mathbf{x},t)$, and consider Fourier modes of \(\delta\phi_k\), with $k$ the comoving momentum. At linear order one finds
\begin{equation}
\ddot{\delta\phi}_k + 3H \dot{\delta\phi}_k
+\left(\frac{k^2}{a^2}+V_{,\phi\phi}(\bar\phi)\right)\delta\phi_k=0\,.
\label{eq:mode_eq}
\end{equation}
Physically, tachyonic amplification arises because the background oscillations periodically drive the effective mass squared $V_{,\phi\phi}(\bar\phi)$ negative. During these intervals the homogeneous condensate temporarily behaves as if it were located near the top of a local hill in the potential, rendering long-wavelength modes \(k^2/a^2<|V_{,\phi\phi}|\) unstable. Unlike parametric resonance \cite{Kofman:1994rk,Kofman:1997yn,Greene:1997fu}, which relies on periodic modulation of a positive mass term, the tachyonic growth is intrinsically explosive, with the instability rate set directly by the curvature scale of the potential \cite{Felder:2000hj,Copeland:2002ku,Felder:2001kt,Bettoni:2019dcw}.
The maximum physical momentum scale involved in tachyonic amplification is determined by the minimum of \(V_{,\phi\phi}\). In the models of interest this yields
\begin{equation}
\frac{k_{\max}}{a}=\sqrt{|V_{,\phi\phi}|_{\min}}
=2^{-3/2}M\simeq 0.35\,M,
\label{eq:kmax}
\end{equation}
with \(M\) the effective mass scale of the intermediate quadratic regime \eqref{eq:chi_c_M}.  Note that this scale does more than just characterizing the linear instability band: it imprints a preferred physical length scale $R_{\text{osc}} \sim M^{-1}$ on the nonlinear structures that subsequently form. As a result, oscillon cores inherit their characteristic size from the curvature of the intermediate quadratic regime. This provides a direct bridge between the effective mass parameter $M$, determined by the gravitational formulation, and the spatial properties of the emergent inhomogeneities.

\section{Nonlinear dynamics, oscillon formation and decay}

The nonlinear stage of preheating in EC HI has been studied with classical lattice simulations  evolving \eqref{eq:KG_full} and \eqref{eq:Friedmann} self-consistently from vacuum-seeded initial conditions \cite{Piani:2023aof,Piani:2025dpy}. A key feature of the dynamics under discussion is the coexistence of disparate time and length scales: the oscillation time scale \(\sim M^{-1}\), the Hubble expansion time \(H^{-1}\), the characteristic instability scale set by \eqref{eq:kmax}, and the long time scales over which oscillons may persist. This motivates a two-pronged numerical strategy.

The first component is a fully nonlinear \(3+1\) dimensional lattice simulation of the scalar field in an expanding background, which captures tachyonic amplification, fragmentation, and oscillon formation. In practice one discretizes the field on a cubic lattice and evolves \eqref{eq:KG_full} together with the Friedmann equation \eqref{eq:Friedmann}, using vacuum initial conditions for the fluctuations. The benchmark simulations are performed on a lattice with $N=288$ points per spatial direction and comoving size $L \simeq 15.7\,M^{-1}$, corresponding to a resolution range $k_{\rm IR} \simeq 0.4\,M$ up to $k_{\rm UV} \simeq 100\,M$. These choices ensure that both the tachyonically amplified modes and the subsequent non-linear structures are well resolved throughout the evolution. The outputs typically include:
(i) the time evolution of \(\langle\phi\rangle\) and \(\delta\phi_{\rm rms}\),
(ii) energy components and the equation of state \(w(t)\),
(iii) power spectra of field fluctuations, and
(iv) spatial maps and histograms of the overdensity \(\Delta\rho\), which provide a direct visual signature of oscillon formation. A limitation of this long-time \(3+1\) simulation in an expanding box is that, as the scale factor grows, fixed comoving resolution corresponds to increasingly coarse physical resolution. While this is not problematic for identifying the onset of fragmentation and the formation of localized lumps, it can become restrictive for accurately following oscillon profiles and their slow energy loss over very long time scales.

This motivates the second component of the analysis: a reduced \(1+1\) dimensional radial simulation aimed at tracking the late-time evolution and lifetime of individual oscillons. Once oscillons are formed and have reached an approximately spherical configuration, their radial profiles can be consistently extracted from the full \(3+1\) classical lattice simulations at a given time slice. These profiles are then used as initial conditions for dedicated \(1+1\) radial simulations, where the assumption of spherical symmetry allows for a substantial increase in spatial resolution and integration time. This approach enables a controlled study of the long-term dynamics, including the gradual emission of radiation, the deformation of the oscillon core, and its eventual decay. In particular, it provides direct access to the oscillon lifetime and the properties of the emitted radiation, which would be prohibitively expensive to resolve within full \(3+1\) simulations.

Together, the \(3+1\) and \(1+1\) approaches provide a coherent picture: the former establishes when and how oscillons form and how much energy they store, while the latter determines their lifetime and the timescale for the onset of radiation domination. This approach is particularly powerful for determining whether oscillons are truly long-lived or whether they decay once their cores probe the quartic regime of the potential. In what follows, we discuss these two stages in detail.

\begin{figure}
  \centering
  \includegraphics[width=\textwidth]{\detokenize{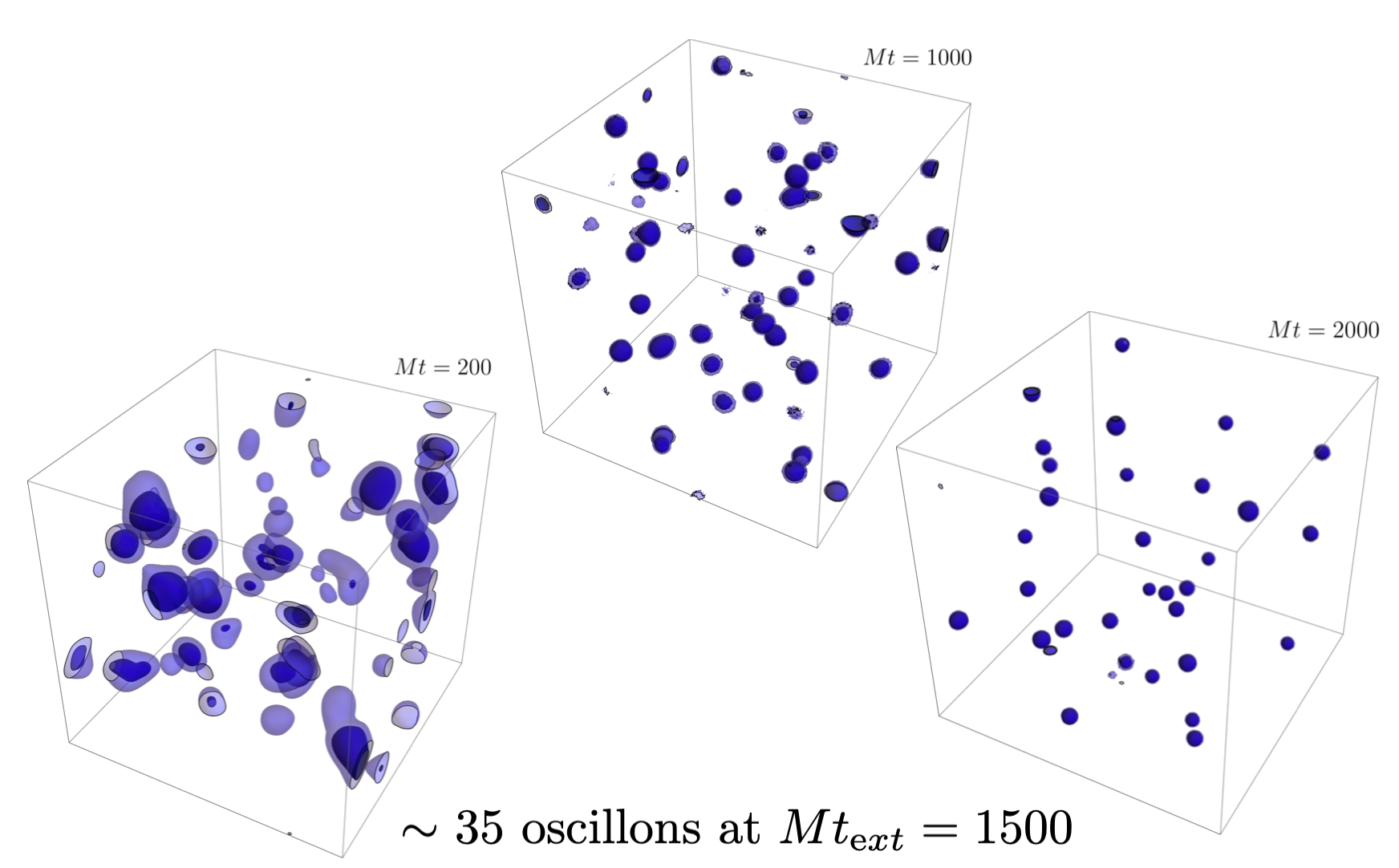}}
  \caption{Three-dimensional visualisation of the inhomogeneous field configuration in lattice simulations at representative times, with $\lambda = 0.001$, $\xi = 5 \cdot 10^4$, and $c = 1.21 \cdot 10^7$. Isosurfaces highlight compact, high-amplitude regions identified as oscillons. As the system evolves, the number and typical size of these objects change due to expansion and gradual radiative losses, ultimately leading to their dissolution and a redistribution of energy into relativistic modes. Adapted from \cite{Piani:2025dpy}.}
  \label{fig:lattice_isosurfaces_oscillons}
\end{figure}

\begin{figure}
  \centering
  \includegraphics[width=0.7\textwidth]{\detokenize{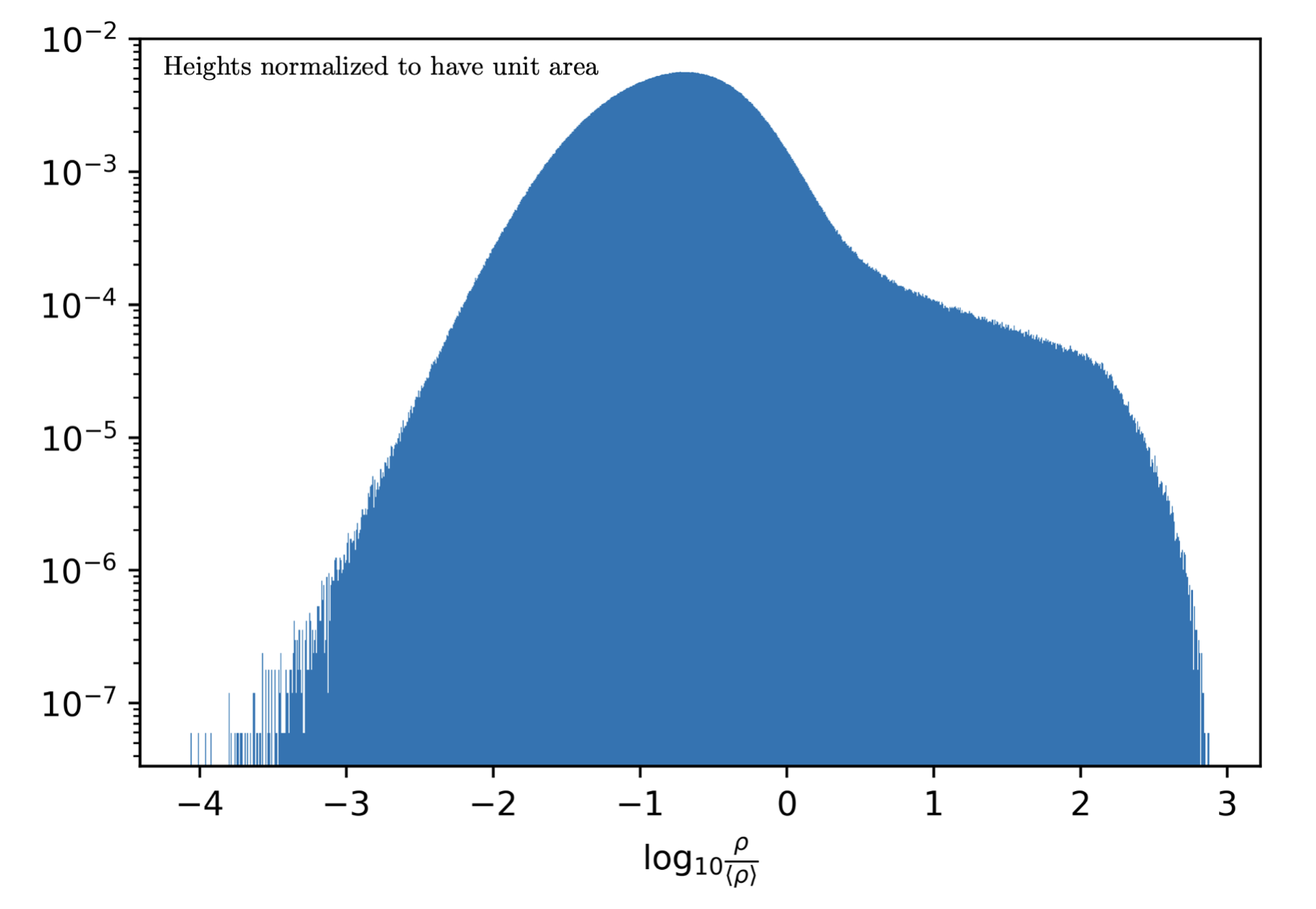}}
  \caption{Distribution of local energy-density contrast during the non-linear stage, shown as a histogram of $\log_{10}(\rho/\bar\rho)$ (normalised to unit area). with $\lambda = 0.001$, $\xi = 5 \cdot 10^4$, and $c = 1.21 \cdot 10^7$. The broad tail towards large overdensities signals the presence of rare, compact regions storing a significant fraction of the total energy---the oscillon population---coexisting with a more dilute background. Adapted from \cite{Piani:2023aof}.}
  \label{fig:rho_histogram}
\end{figure}

\subsection{$3+1$ classical lattice simulations}

As tachyonic amplification proceeds, the homogeneous condensate fragments into localized overdensities. In the intermediate EC regime, a significant fraction of these overdensities organize into oscillon configurations: localized, quasi-periodic lumps that can be remarkably long-lived compared to the oscillation time scale. In particular, localized overdensities characteristic of oscillon configurations are already clearly identifiable at times $Mt \sim \mathcal{O}(10^2)$, shortly after the end of the tachyonic amplification stage. These structures are associated with a pronounced peak in the scalar power spectrum located at
${k}/{(aM)} \simeq 0.5$,  which tracks the typical inverse size of the emergent configurations.

\begin{figure}
  \centering
  \includegraphics[width=0.98\textwidth]{\detokenize{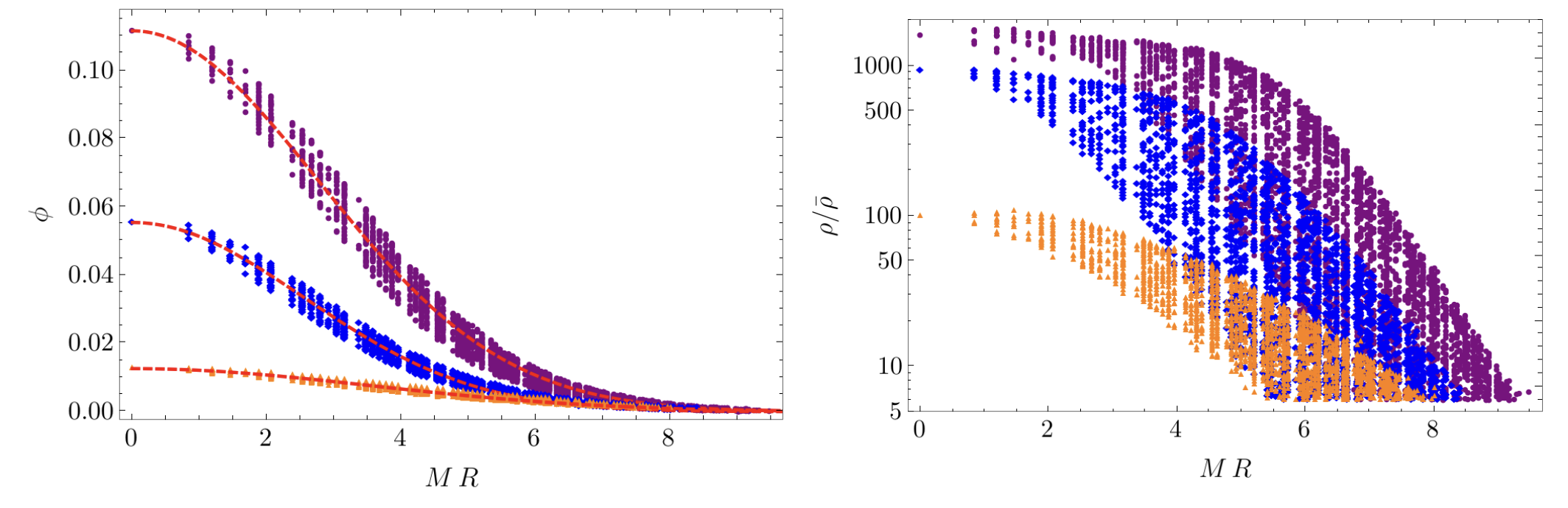}}
  \caption{Characterisation of oscillon profiles extracted from lattice simulations with $\lambda = 0.001$, $\xi = 5 \cdot 10^4$, and $c = 1.21 \cdot 10^7$. \textit{Left:} Field amplitude $\phi$ as a function of the rescaled radius $MR$, for representative oscillon realisations (markers) together with smooth fits (dashed lines). \textit{Right:} Corresponding central overdensity $\rho/\bar\rho$ versus $MR$, highlighting the correlation between compactness and energy concentration. These diagnostics are used to build effective initial data for $1+1$ dimensional simulations tracking the long-time evolution and decay of individual oscillons. Adapted from \cite{Piani:2025dpy}.}
  \label{fig:oscillon_profiles_MR}
\end{figure}

A practical way to identify oscillons in simulations is to study the energy overdensity
\begin{equation}
\Delta\rho(\mathbf{x},t)\equiv \frac{\rho(\mathbf{x},t)}{\bar\rho(t)},
\qquad
\rho=\frac12\dot\phi^2+\frac{(\nabla\phi)^2}{2a^2}+V(\phi),
\end{equation}
and to select localized regions with \(\Delta\rho\) above a threshold (e.g.\ \(\Delta\rho>5\) or \(\Delta\rho>10\)), tracking them in time. The emergence of localized high-density regions in the lattice simulations is visualised in Fig.~\ref{fig:lattice_isosurfaces_oscillons}, while Fig.~\ref{fig:rho_histogram} shows the corresponding distribution of energy-density contrasts, exhibiting a broad tail associated with the oscillon population. Once such regions are identified, their spatial profiles are typically close to spherically symmetric and can be characterized with simple fitting functions. A useful phenomenological parametrization for the field is a Gaussian profile with slowly varying amplitude \(A(t)\) and radius \(R(t)\),
\begin{equation}
\phi(t,r)\;\approx\; A(t)\,e^{-r^2/R(t)^2}\cos(\omega t)\,,
\label{eq:osc_profile}
\end{equation}
which describes well the cores extracted from lattice data and provides an intuitive description of their subsequent evolution. Representative field profiles extracted from the lattice are shown in Fig.~\ref{fig:oscillon_profiles_MR}, where the radial dependence of the oscillon core is compared with the previous smooth fit, illustrating the approximate spherical symmetry and coherence of the configurations at extraction time. 

\begin{figure}
  \centering
  \includegraphics[width=0.8\textwidth]{\detokenize{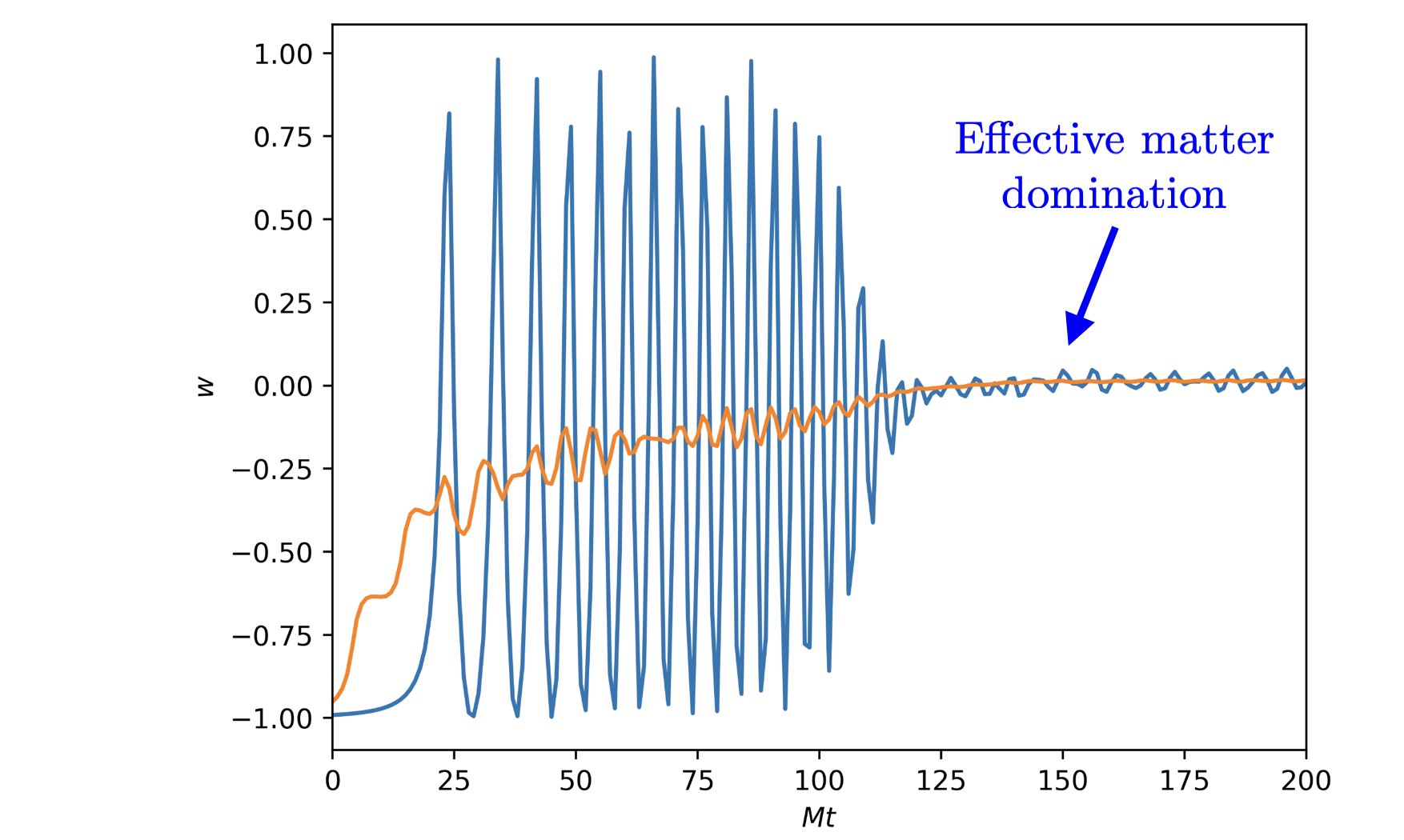}}
  \caption{Time evolution of the equation-of-state parameter $w \equiv \langle p\rangle/\langle\rho\rangle$ in the aftermath of inflation, with $\lambda = 0.001$, $\xi = 5 \cdot 10^4$, and $c = 1.21 \cdot 10^7$. Large oscillations and sudden damping reflect respectively coherent inflaton dynamics and violent fragmentation, while the drift of the averaged equation of state towards $\bar w\simeq 0$ indicates a stage with effective matter-like behaviour driven by an oscillon-dominated energy budget. At later times, radiative losses and oscillon decay push the system towards $\bar w\simeq 1/3$, signalling the onset of radiation domination. Adapted from \cite{Piani:2023aof}.}
  \label{fig:eos_evolution}
\end{figure}

To quantify the impact of oscillons on the expansion history, one can define the fraction of energy stored in oscillons by summing the energy contained in the identified localized regions and dividing by the total energy,
\begin{equation}
f_{\rm osc}(t)\equiv \frac{\rho_{\rm osc}(t)}{\bar\rho(t)}\,.
\label{eq:fosc}
\end{equation}
For suitable parameters, simulations show that \(f_{\rm osc}\) can become \(\mathcal{O}(0.5\text{--}0.7)\) shortly after formation, and the volume-averaged equation of state,
\begin{equation}
w(t)\equiv \frac{\bar p}{\bar\rho},
\qquad
\bar\rho=\left\langle \frac12\dot\phi^2+\frac{(\nabla\phi)^2}{2a^2}+V(\phi)\right\rangle,
\qquad
\bar p=\left\langle \frac12\dot\phi^2-\frac{(\nabla\phi)^2}{6a^2}-V(\phi)\right\rangle,
\label{eq:w_def}
\end{equation}
approaches matter-like behavior, \(w\simeq 0\), over the oscillon-dominated stage. This provides a natural mechanism for a transient matter-like phase before radiation domination. Importantly, even when the average equation of state is close to zero, the system is highly inhomogeneous, and the subsequent decay of oscillons can lead to a relatively sudden change in the effective equation of state. The time evolution of the averaged equation-of-state parameter is displayed in Fig.~\ref{fig:eos_evolution}, where large oscillations signal coherent inflaton dynamics and fragmentation, followed by a transient matter-like stage before the approach to radiation domination.

\begin{figure}
  \centering
  \includegraphics[width=0.7\textwidth]{\detokenize{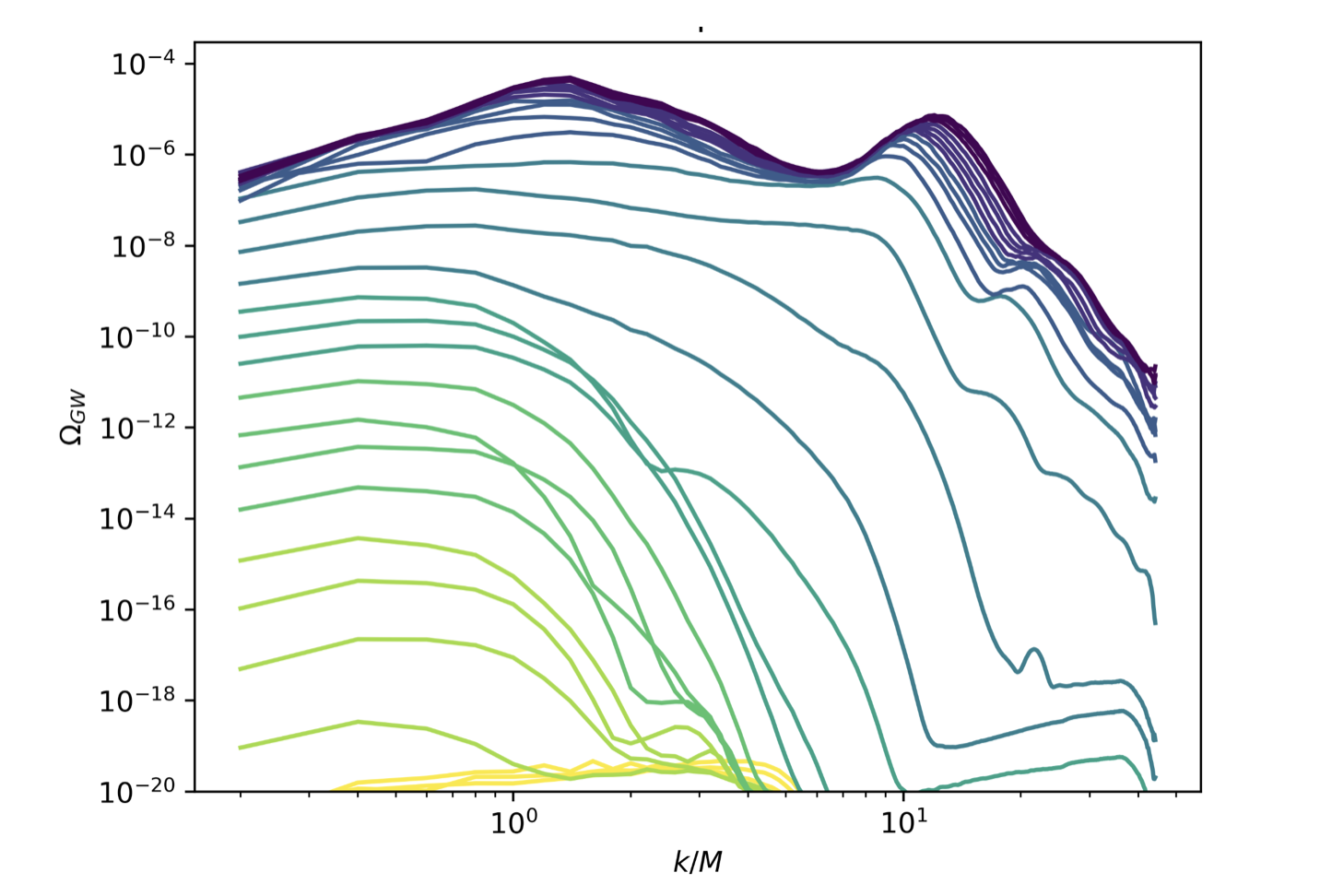}}
  \caption{Time evolution of the gravitational-wave spectral energy density $\Omega_{\rm GW}(k)$ as a function of the rescaled momentum $k/M$, with $\lambda = 0.001$, $\xi = 5 \cdot 10^4$, and $c = 1.21 \cdot 10^7$. The spectrum builds up during fragmentation and oscillon formation, typically developing a peak at momenta corresponding to the characteristic size of the dominant inhomogeneities, and evolves as the system approaches radiation domination. Although the associated present-day frequencies are generally ultra-high in Higgs-inflation scenarios, the signal constitutes a robust prediction of the non-linear dynamics. Adapted from \cite{Piani:2023aof}.}
  \label{fig:GW_spectrum_evolution}
\end{figure}

The violent fragmentation and subsequent oscillon formation source gravitational waves through the anisotropic stress of the scalar field. As the system evolves and the oscillons become approximately spherical, the emission becomes increasingly inefficient due to the suppression of the quadrupole moment, and the GW spectrum effectively freezes. The signal is commonly presented as a spectral energy density,
\begin{equation}
\Omega_{\rm GW}(k,t)\equiv \frac{1}{\rho_c(t)}\frac{d\rho_{\rm GW}}{d\ln k}\,,
\qquad \rho_c(t)=3M_P^2H^2(t)\,.
\label{eq:OmegaGW_def}
\end{equation}
Redshifting \eqref{eq:OmegaGW_def} to the present epoch can be expressed schematically as
\begin{equation}
\Omega_{\rm GW,0}(k)\simeq \Omega_{\rm GW}(k,t_{\rm gen})
\left(\frac{a_{\rm gen}}{a_0}\right)^4
\left(\frac{H_{\rm gen}}{H_0}\right)^2\,,
\label{eq:OmegaGW_redshift}
\end{equation}
with the peak frequency set by the characteristic physical scale of the source at generation. If the peak momentum at production is \(k_{\rm p}\sim a_{\rm gen}/R_{\rm osc}\), with \(R_{\rm osc}\sim \alpha/M\) and \(\alpha\) an order-one factor extracted from simulations, then the present-day peak frequency is
\[
f_0 \sim \frac{k_{\rm p}}{2\pi a_0}
\sim \frac{1}{2\pi}\,\frac{a_{\rm gen}}{a_0}\,\frac{M}{\alpha}.
\]
The build-up of the gravitational-wave spectrum during fragmentation and oscillon dynamics is illustrated in Fig.~\ref{fig:GW_spectrum_evolution}, which shows the time evolution of $\Omega_{\rm GW}(k)$ and the emergence of a peak associated with the characteristic oscillon scale, \(R\sim M^{-1}\). In Higgs-inflation realizations, the characteristic frequencies are typically ultra-high, far outside the sensitivity range of current and planned detectors, but the signal provides a robust and calculable prediction of the nonlinear dynamics. 

Overall, these results suggest that oscillons can play a major role in the early post-inflationary history. However, their ultimate fate is controlled by the small-field structure of the Higgs potential, as we discuss next using $1+1$ radial lattice simulations.

\subsection{$1+1$ radial lattice simulations}

\begin{figure}
  \centering
  \includegraphics[width=0.8\textwidth]{\detokenize{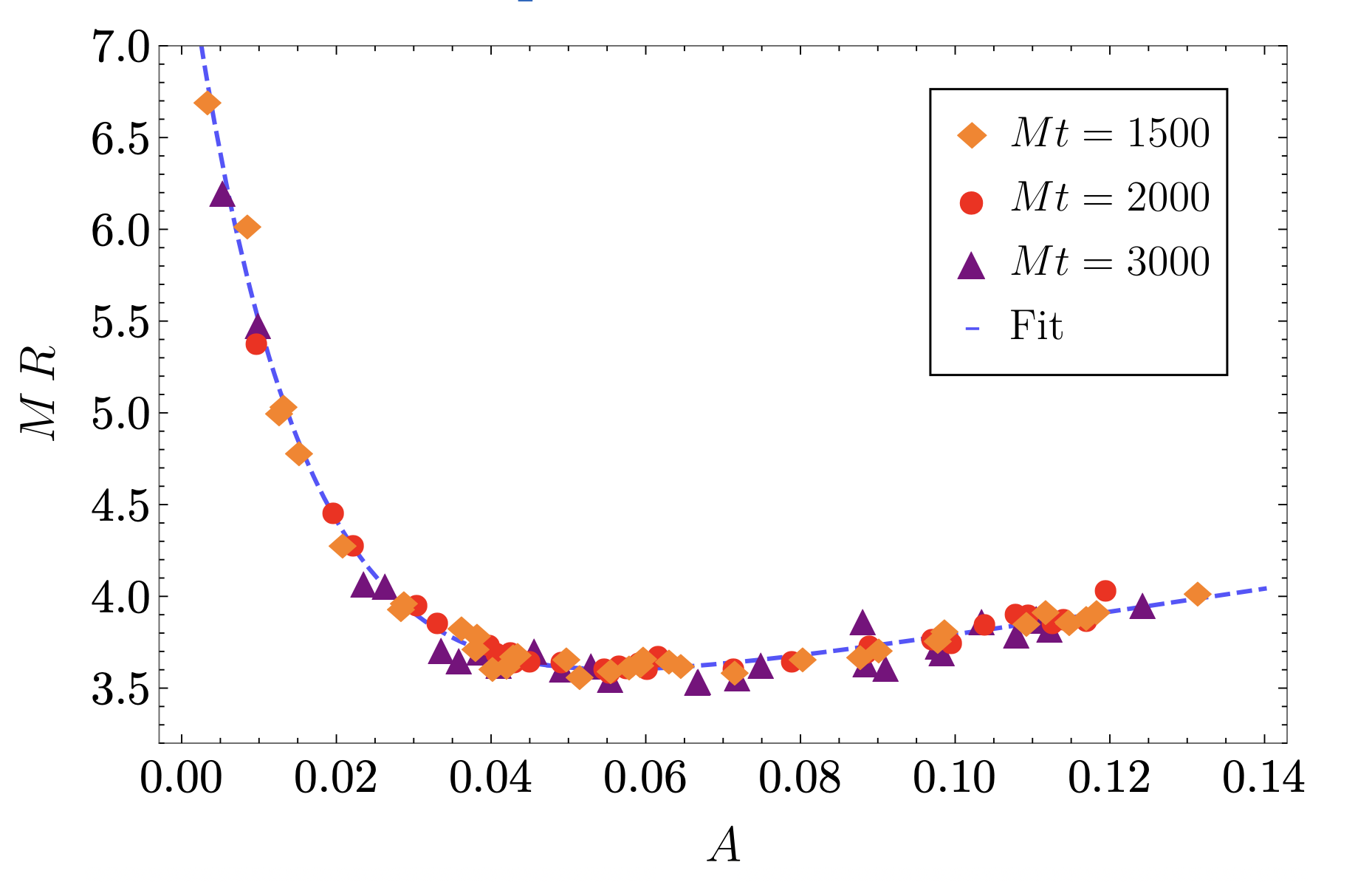}}
  \caption{Correlation between the oscillon core amplitude parameter $A$ and the characteristic radius $R$ for profiles extracted at different times, with $\lambda = 0.001$, $\xi = 5 \cdot 10^4$, and $c = 1.21 \cdot 10^7$. The dashed curve shows a phenomenological fit capturing the observed $R(A)$ relation across the oscillon population. This relation provides a compact summary of the family of localized configurations produced during the non-linear stage and is used to initialise $1+1$ dimensional simulations of individual oscillon evolution. Adapted from \cite{Piani:2025dpy}.}
  \label{fig:R_vs_A_timeslices_fit}
\end{figure}

A crucial feature of the present scenario is that the effective potential  \eqref{eq:V_piecewise} interpolates between a quadratic regime at intermediate field values and a quartic regime close to the origin.  
 Oscillons are supported as long as their core amplitude probes an effectively quadratic region of the potential. While $A(t) \gtrsim \phi_c$, the core oscillates in a regime where radiation into linear modes is suppressed, and the configuration behaves as a quasi-stable localized lump. However, in contrast to standard oscillon scenarios, radiation is continuously emitted from the outer layers, where the field amplitude remains below the crossover scale and continuously probes the quartic regime. This induces a persistent leakage of energy that drives the configuration towards smaller amplitudes and accelerates its decay. As a result, the amplitude gradually decreases and eventually satisfies $A(t_d) \sim \phi_c$.  Once this threshold is reached, the quartic self-interactions become dynamically relevant, efficient mode coupling sets in, and the configuration rapidly dissolves into relativistic scalar waves. In this sense, Higgs-like oscillons in the EC framework are generically ephemeral: the quartic–quadratic transition enforces a built-in dynamical cutoff on their lifetime.

A complementary way to visualize this mechanism is provided by the evolution in the $(A,MR)$ plane. Oscillons extracted from the $3\!+\!1$ lattice evolve approximately along an empirical one-parameter family $R(A)$, forming a quasi-adiabatic manifold of localized solutions, cf.~Fig.~\ref{fig:R_vs_A_timeslices_fit}. The decay can then be interpreted as the endpoint of this trajectory: as the amplitude decreases and the configuration drifts toward smaller $A$, it eventually reaches a region of parameter space where no long-lived localized solutions exist. At that point energy loss accelerates and the core disperses.

Typical lifetimes extracted from the radial simulations satisfy $
M(t_d - t_{\rm ext}) \sim \mathcal{O}(10^3\text{--}10^4)$,  much longer than the oscillation time scale $M^{-1}$, yet parametrically finite. Importantly, the lifetime does not grow arbitrarily with the underlying parameters: the inevitable entry of the oscillon core into the quartic regime prevents the emergence of exponentially long-lived configurations.

The radial $1+1$ lattice simulations provide a clean quantitative handle on this process. Starting from profiles extracted at time $t_{\rm ext}$, one evolves them forward in a high-resolution spherical setup and monitors the oscillon energy 
\begin{equation}
E_{\rm osc}(t)=4\pi\int_0^\infty dr\,r^2\left[
\frac12\dot\phi^2+\frac12(\partial_r\phi)^2+V(\phi)
\right],
\label{eq:Eosc}
\end{equation}
as well as the evolution of fitted profile parameters \(A(t)\) and \(R(t)\) in \eqref{eq:osc_profile}. 
\begin{figure}
  \centering
  \includegraphics[width=0.8\textwidth]{\detokenize{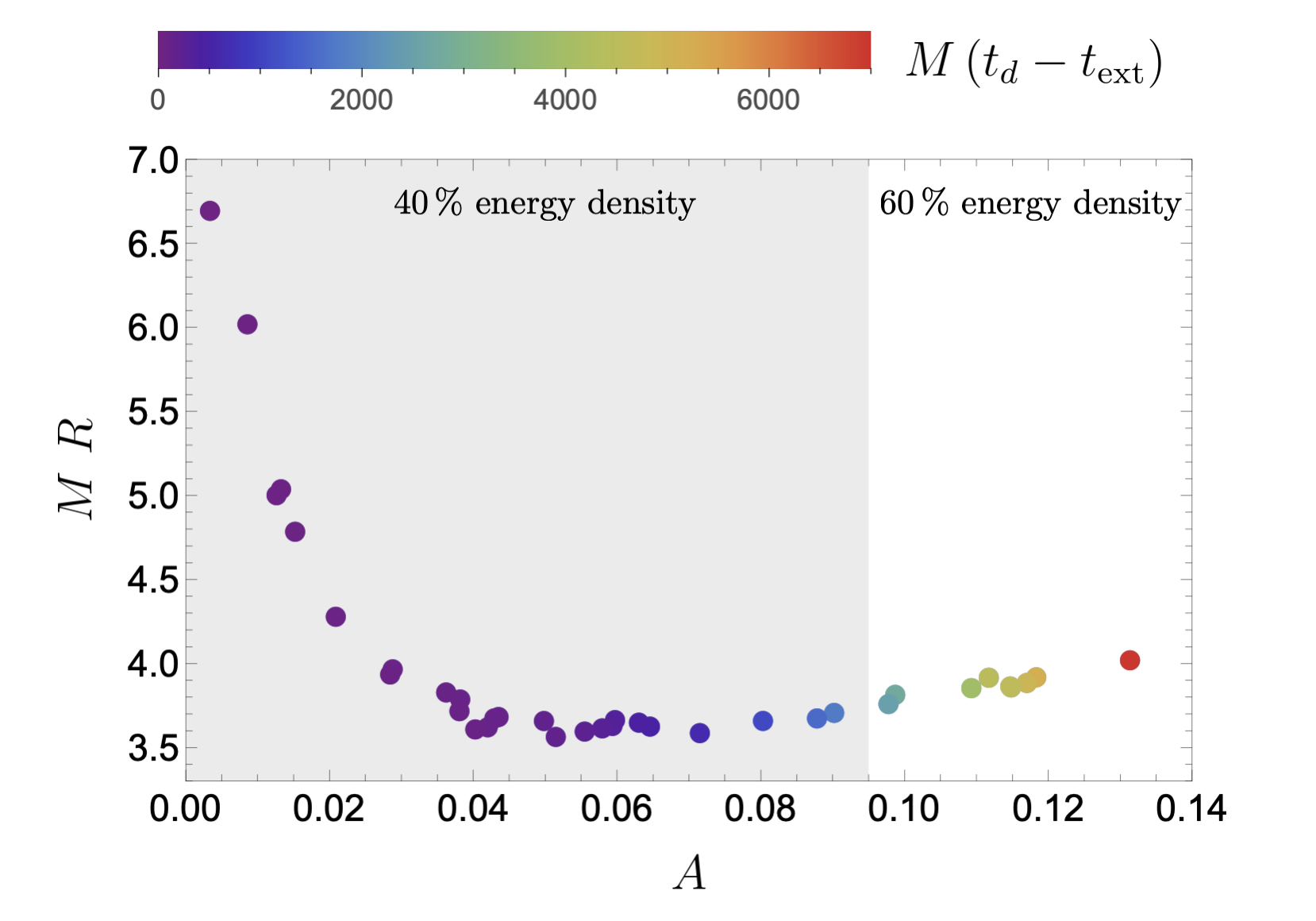}}
  \caption{Time to decay for representative oscillons as a function of their extracted parameters $(A,MR)$, with $\lambda = 0.001$, $\xi = 5 \cdot 10^4$, and $c = 1.21 \cdot 10^7$. The colour bar indicates the lifetime measured from the extraction time, $M(t_d-t_{\rm ext})$. Shaded regions illustrate subsamples contributing different fractions of the total energy density at extraction, highlighting that the most energetic (and typically most compact) configurations dominate the energy budget and therefore control the duration of the oscillon-dominated stage. Adapted from \cite{Piani:2025dpy}.}
  \label{fig:oscillon_lifetimes_energy_bins}
\end{figure}
\begin{figure}
  \centering
  \includegraphics[width=0.75\textwidth]{\detokenize{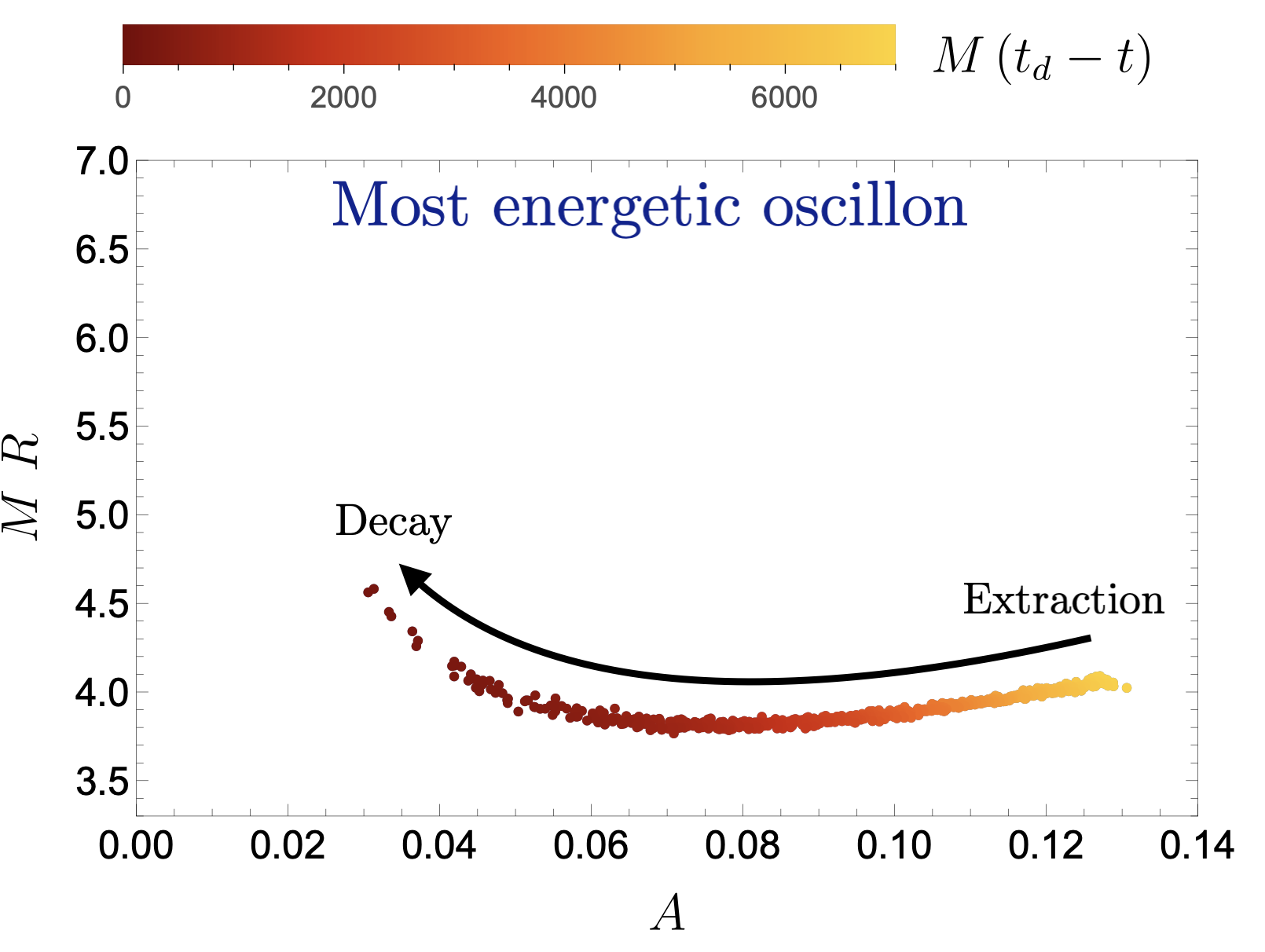}}
  \caption{Evolution of the most energetic oscillon in the $(A,MR)$ plane after it is identified in the $3\!+\!1$ lattice and evolved in a high-resolution spherical setup, with $\lambda = 0.001$, $\xi = 5 \cdot 10^4$, and $c = 1.21 \cdot 10^7$. Colour encodes the time remaining to decay, $M(t_d-t)$, illustrating the drift along the empirical $R(A)$ manifold toward the decay region. Adapted from \cite{Piani:2025dpy}.}
  \label{fig:most_energetic_oscillon_trajectory}
\end{figure}
As shown in Fig.~\ref{fig:oscillon_lifetimes_energy_bins}, the most energetic (and typically most compact) oscillons dominate the total energy stored in the localized population even if they are subdominant in number. Their decay therefore controls, to a large extent, the time at which the oscillon-dominated stage ends. The trajectory of the most energetic oscillon is illustrated in Fig.~\ref{fig:most_energetic_oscillon_trajectory}. As the amplitude decreases and the configuration approaches the quartic regime, the profile broadens, scalar radiation becomes more efficient, and the remaining lifetime shortens rapidly. The drift along the $R(A)$ manifold provides a direct visualization of gradual energy loss and the approach to dissolution. Quantitatively, the lifetime of individual oscillons is found to satisfy
$M\,(t_{\rm dec}-t_{\rm ext}) \sim \mathcal{O}(10^3\!-\!10^4)$, where $t_{\rm ext}$ denotes the time at which the oscillon is extracted from the lattice simulation.

From a cosmological perspective, the key point is that the oscillon stage is necessarily transient. Although oscillons can temporarily store a substantial fraction of the total energy density and drive the averaged equation of state toward $\bar w \simeq 0$, their finite lifetime prevents the generation of an arbitrarily prolonged matter-dominated epoch. The decay of the dominant oscillons triggers a relatively rapid transfer of energy into relativistic modes, pushing the system toward radiation domination.

\begin{figure}
  \centering
  \includegraphics[width=0.8\textwidth]{\detokenize{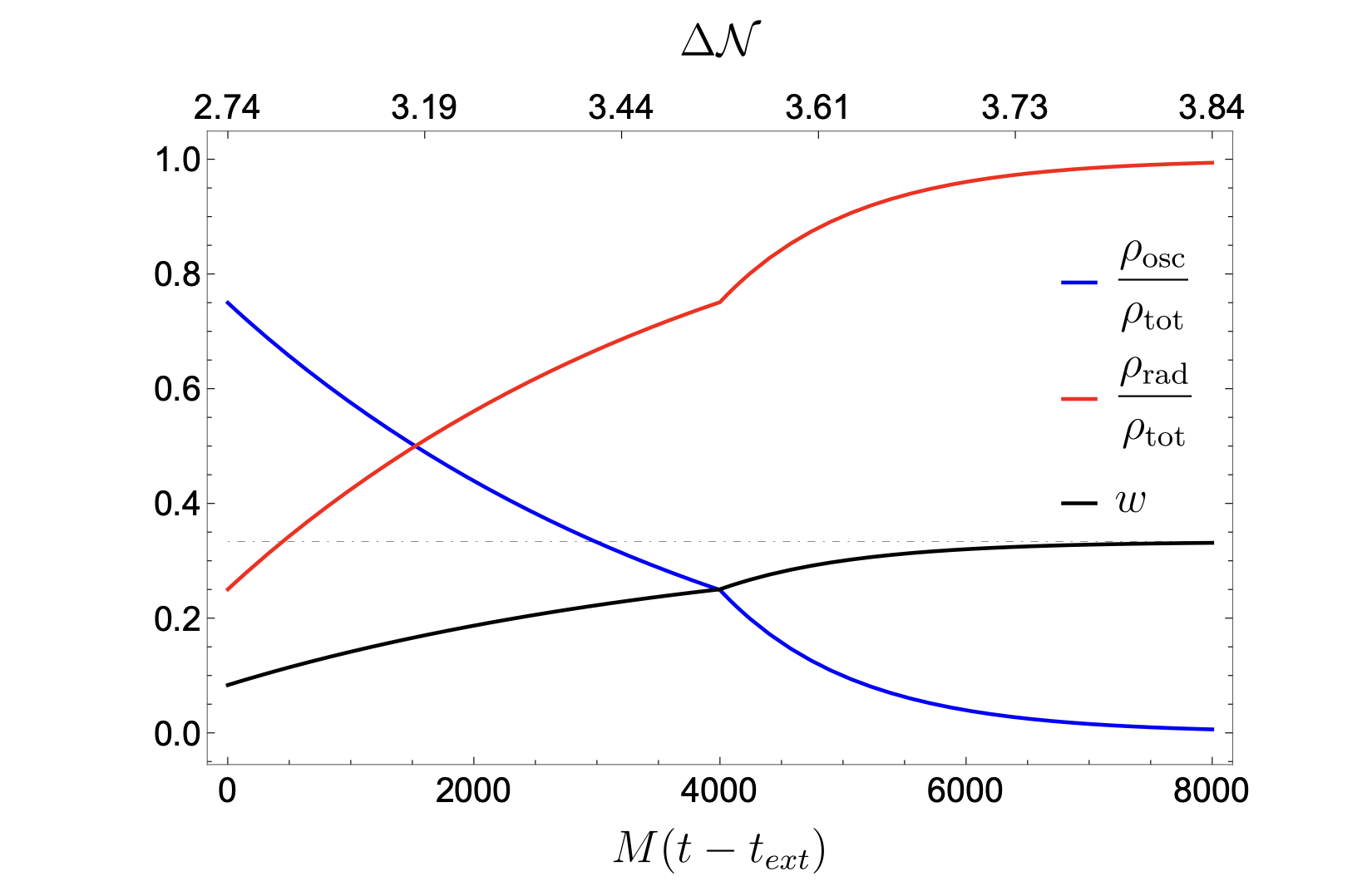}}
  \caption{Late-time evolution of the energy budget after the oscillon population is identified at $t_{\rm ext}$, with $\lambda = 0.001$, $\xi = 5 \cdot 10^4$, and $c = 1.21 \cdot 10^7$. The blue curve shows the oscillon energy fraction $\rho_{\rm osc}/\rho_{\rm tot}$, while the red curve shows the complementary fraction in relativistic modes, $\rho_{\rm rad}/\rho_{\rm tot}$. The black curve displays the corresponding averaged equation-of-state parameter $\bar w$. The gradual depletion of the oscillon component and the rise of the relativistic component drive the transition toward radiation domination and quantify the finite duration of the intermediate matter-like stage. Taken from \cite{Piani:2025dpy}.}
  \label{fig:energy_budget_and_w}
\end{figure}

The transfer of energy from the oscillon population into relativistic modes is explicitly shown in Fig.~\ref{fig:energy_budget_and_w}. The depletion of $\rho_{\rm osc}$ and the corresponding rise of $\rho_{\rm rad}$ drive the averaged equation of state away from $\bar w \simeq 0$ toward $\bar w \simeq 1/3$. The finite lifetime of oscillons therefore bounds the duration of the heating phase and stabilizes the post-inflationary expansion history in this class of models.

\section{Cosmological implications}

The finite duration of the oscillon-dominated phase impacts the expansion history between the end of inflation and the onset of radiation domination, and therefore the inferred value of \({\cal N}_*\) entering \eqref{eq:infl_obs}. In particular, a prolonged matter-like stage (\(\bar w\simeq 0\)) would shift \({\cal N}_*\) more strongly than a radiation-like stage (\(\bar w\simeq 1/3\)). Since the spectral tilt and tensor-to-scalar ratio scale as $n_s-1\propto 1/{\cal N}_*$  and \(r\propto {\cal N}_*^{-2}\), bounding the duration of the matter-like stage stabilizes the predictions for \((n_s,r)\) against large reheating uncertainties. 

A more explicit relation can be written by matching the pivot scale to the present horizon through the expansion history. Denoting by \(\rho_{\rm end}\) the energy density at the end of inflation and by \(\rho_{\rm reh}\) the energy density at the onset of radiation domination,  the number of $e$-folds between horizon exit and the end of inflation takes the form
\begin{equation}
{\cal N}_* = 61.5 + \frac{1}{4}\ln\!\left(\frac{V_*^2}{M_P^4 \rho_{\rm end}}\right)
+ \frac{3\bar w -1}{4}\,\Delta {\cal N}_{\rm reh}\,,
\end{equation}
where $\bar w$ denotes the effective equation of state during the post-inflationary stage and
\begin{equation}
\Delta {\cal N}_{\rm reh}\equiv \ln\!\left(\frac{a_{\rm reh}}{a_{\rm end}}\right)
=
\frac{1}{3(1+\bar w)}\ln\!\left(\frac{\rho_{\rm end}}{\rho_{\rm reh}}\right)\,.
\end{equation}
In practical terms, this means that the reheating uncertainty in ${\cal N}_*$ is dynamically bounded from above. Even though oscillons can temporarily drive $\bar w \simeq 0$, their finite lifetime prevents an arbitrarily long matter-dominated phase, with $\Delta \mathcal{N}_{\rm end}=3.81$ $e$-folds. Consequently, the shift in ${\cal N}_*$ induced by reheating remains moderate, and the HI inflationary predictions \eqref{eq:infl_obs} retain a degree of robustness that would otherwise be lost in scenarios with prolonged oscillon domination.

A self-consistent way to incorporate the impact of the post-inflationary dynamics in the inflationary observables is to proceed iteratively. One starts from a fiducial value of ${\cal N}_*$ to fix the parameters of the inflationary potential, then follows the subsequent evolution through the oscillon-forming stage and the eventual onset of radiation domination, and finally extracts an updated value of ${\cal N}_*$ from the resulting expansion history. Repeating this procedure would in principle lead to convergence once the input and output values of ${\cal N}_*$ agree within the desired accuracy. For the benchmark configuration considered here, $\{\lambda,\xi,c\}=\{0.001,\,5\times10^4,\,1.21\times10^7\}$, the fiducial choice gives an initial estimate ${\cal N}_*^{(0)}=55$. In practice, however, going beyond the first iteration is not entirely straightforward, since updating the model parameters may alter the post-inflationary evolution itself by modifying the relative hierarchy between the characteristic field scales $(\phi_{\rm end},\phi_i,\phi_c)$. For this reason, we restrict ourselves to the first refinement. Combining the information from the $3+1$ lattice simulations with the late-time $1+1$ radial analysis, we obtain
\begin{equation}
\label{eq:observables-corrected}
{\cal N}_*^{(1)}\simeq 53\,,\qquad \qquad 
n_s \simeq 1-\frac{2}{{\cal N}_*^{(1)}} \simeq 0.9623\,.
\end{equation}

\section{Conclusions and outlook}

We have reviewed the preheating dynamics of HI in EC gravity, focusing on the interplay between tachyonic instabilities, nonlinear fragmentation, and oscillon dynamics. In this framework, integrating out non-propagating torsion selects a predictive scalar-tensor dynamics in the Einstein frame, characterized by a Higgs-like potential with a quartic regime at small field values, an intermediate quadratic region, and an inflationary plateau at large field values. This structure leads to a distinctive heating phenomenology in which tachyonic amplification can efficiently seed condensate fragmentation and oscillon formation in a controlled intermediate regime.

A central point is that, while oscillons can temporarily store a substantial fraction of the energy density and induce an intermediate matter-like expansion stage, their fate is ultimately governed by the small-field behavior of the potential. As the oscillon cores drift into the quartic regime, self-interactions efficiently radiate energy into relativistic modes, rendering these objects short-lived. The resulting decay drives a comparatively rapid onset of radiation domination and bounds the duration of the heating phase. This, in turn, stabilizes the mapping between inflationary dynamics and late-time observables by restricting the range of post-inflationary expansion histories compatible with the model.

Overall, EC HI provides a concrete example in which the gravitational formulation leaves scalar perturbations essentially intact while radically reshaping the nonlinear reheating epoch. The emergence and eventual decay of oscillons are not accidental features, but controlled consequences of the quartic--quadratic--plateau structure of the Einstein-frame potential. Preheating thus becomes a sensitive probe of the underlying gravitational and EFT structure, illustrating that the path from inflation to the hot Big Bang carries detailed information about how gravity is formulated at high energies.

Several directions remain open. Incorporating the full Standard Model field content in nonlinear simulations would sharpen the quantitative picture and assess the robustness of the oscillon stage in the presence of gauge fields and fermions. It would also be interesting to extend the analysis to broader classes of scalar-tensor theories arising from other gravitational formulations, and to explore how the associated ultra-high-frequency gravitational-wave spectra might serve as a probe of the microphysics of preheating.

\section{Acknowledgments}

J.~R. is supported by a Ram\'on y Cajal contract of the Spanish Ministry of Science and Innovation with Ref.~RYC2020-028870-I. This research was further supported by the project PID2022-139841NB-I00 of MICIU/AEI/10.13039/501100011033 and FEDER, UE. 

\bibliographystyle{JHEP}
\bibliography{bibliography}
\end{document}